\documentstyle[11pt,epsfig,cite]{article} 
\def\baselinestretch{1.3}
\catcode`@=11 
\def \gsim{\mathrel{\mathpalette\@versim>}}
\def \lsim{\mathrel{\mathpalette\@versim<}}
\def \@versim#1#2{\lower0.4ex\vbox{\baselineskip\z@skip\lineskip\z@skip
     \lineskiplimit\z@\ialign{$\m@th#1\hfil##\hfil$%
     \crcr#2\crcr\sim\crcr}}}
\catcode`@=12 
\newcommand{\ba}{\begin{array}}
\newcommand{\ea}{\end{array}}
\newcommand{\bd}{\begin{displaymath}}
\newcommand{\ed}{\end{displaymath}}
\newcommand{\be}{\begin{equation}}
\newcommand{\ee}{\end{equation}}
\newcommand{\bea}{\begin{eqnarray}}
\newcommand{\eea}{\end{eqnarray}}
\def\dis{\displaystyle}

\begin{document}
\begin{flushright}
{\large  MRI-P-020201 \\
HIP-2002-16/TH \\
May, 2002  \\ 
hep-ph/0205103}
\end{flushright}

\begin{center}
{\Large\bf Radiative production of invisible charginos in 
        photon-photon collisions}\\[20mm]
Debajyoti Choudhury \footnote{E-mail: debchou@mri.ernet.in},  
Biswarup Mukhopadhyaya \footnote{E-mail: biswarup@mri.ernet.in}
and Subhendu Rakshit \footnote{E-mail: srakshit@mri.ernet.in} \\
{\em Harish-Chandra Research Institute,\\
Chhatnag Road, Jhusi, Allahabad - 211 019, India}. \\
and \\
Anindya Datta \footnote{E-mail: datta@pcu.helsinki.fi} \\
{\em Helsinki Institute of Physics,\\
University of Helsinki, P.O. Box 64, Helsinki, Finland}. \\
\end{center}
\begin{abstract}
  If in a supersymmetric model, the lightest chargino is nearly
  degenerate with the lightest neutralino, the former can decay
  into the latter alongwith a soft pion (or a lepton-neutrino
  pair). Near degeneracy of the chargino and neutralino masses can cause
  the other decay products (the pion or the lepton) to be almost invisible. 
  Photon-photon colliders offer a possibility of clean
  detection of such an event through a hard photon tag.
\end{abstract}

\vskip 1 true cm

\newpage
\setcounter{footnote}{0}

\def\baselinestretch{1.8}

\section{Introduction}

Supersymmetry (SUSY) is widely recognized as a possibility in our
quest for physics beyond the Standard Model.  The particle spectrum of
a particular SUSY model depends on the dynamics of the SUSY breaking.
Determination of the SUSY breaking mechanism thus constitutes an
integral part of any new accelerator proposal.  Different sectors of
the supersymmetric model can often be independently investigated in
such contexts.  An important component of such a model is the
chargino-neutralino sector, where the physical states are formed as
linear superpositions of the charged (or neutral, as the case may be)
gauginos and Higgsinos respectively.  A clear and unambiguous
observation of this sector is important, as this will not only tell us
about the SUSY breaking parameters lending themselves as gaugino
masses, but also reveal crucial details of the mixing process
operative here, driven by quantities such as the Higgsino mass $\mu$
and the ratio of the vacuum expectation values of the two Higgs
doublets present in the model.

Direct searches at the Large Electron Positron (LEP) collider have
already set lower limits of about 99 GeV~\cite{lep_limit} on the mass
of the lightest chargino. Charginos and neutralinos with higher masses
can be explored at Run II of the Tevatron or at the Large Hadron
Collider (LHC), where a useful signal arises from the `hadronically
quiet' trilepton final states formed by decays of the $\chi_2^0
\chi_1^{\pm}$ pair~\cite{clean_tri_lep}. Here $\chi_1^{\pm}$ is the
lightest chargino and $\chi_2^0$, the next-to-lightest neutralino.
However, this mode becomes hard to tag to whenever the leptonic decay
channels of the $\chi_1^{\pm}$ get suppressed. This can happen, for
example, in theories with anomaly mediated SUSY breaking
(AMSB)~\cite{AMSB}.  Anomaly mediated models attempt to link the
dynamics of SUSY breaking to models with extra compactified
dimensions. The SUSY breaking sector is confined to a 3-brane
separated from the one on which the standard model fields reside. SUSY
breaking is conveyed to the observable sector by a super-Weyl anomaly.
One remarkable feature of these models is the proportionality of the
soft gaugino masses to the corresponding gauge beta-function
coefficient ($b_i$). Since $b_2$ (corresponding to $SU(2)$) is smaller
than $b_1$ (same for $U(1)_Y$), the $SU(2)$ gaugino mass $M_2$ is
smaller than $M_1$.  As a result, the lightest neutralino
($\chi_1^0$), the LSP, is practically degenerate with $\chi_1^{\pm}$
and both are wino-like.  With the mass separation being a few hundred
MeVs at best, the $\chi_1^{\pm}$ decays dominantly into a $\chi_1^0$
(which is stable and invisible so long as lepton and baryon numbers
are conserved) and a soft pion. The lower limit on the mass of such a
chargino is around 87 GeV~\cite{lep_limit}. Thus a $\chi_1^0
\chi_1^{\pm}$ pair, as opposed to $\chi_2^0 \chi_1^{\pm}$ whose
production rate is suppressed, essentially escapes undetected. One way
out here is to look for macroscopic tracks left by the
chargino~\cite{macro_track}, but the success of this strategy is not
guaranteed. Alternative signals for such a scenario have been proposed
and studied in detail in the context of a high-energy
electron-positron collider, {\it via} an analysis of the `single
photon plus missing energy' signals arising from $e^{+}e^{-}
\longrightarrow \chi_1^{+} \chi_1^{-} \gamma$~\cite{chen_drees,
  single_photon, single_photon_2}.  In this paper, we suggest another
possibility, namely, the radiative production of chargino pairs in
photon-photon collision, triggered by laser back-scattering in a
linear $e^{+}e^{-}$ or $e^{-}e^{-}$ collider.

The advantage of the photon-photon collision mode is that the
production is controlled only by electromagnetic interaction.  Thus
the chargino production rates are independent of the mixing mechanism.
Also, in an electron-positron collider the single photon signals are
plagued with backgrounds from $e^{+}e^{-} \longrightarrow \nu
\bar{\nu} \gamma$.  A similar background is virtually nonexistent in
photon-photon collisions. Since only charged particles are formed {\it
  via} $\gamma\gamma$ collisions at the tree level, the single photon
visible final states do not get any contribution from $\chi^0$ pairs,
unlike in the case of $e^+e^-$ collisions. And lastly, as we shall
discuss in further detail in section 4, the suggested signals in an
$e^{+}e^{-}$ collider can be strongly affected by the sneutrino
mass~\cite{single_photon}, at least in a very significant region of
the parameter space. Such model dependence is completely avoided in
the $\gamma\gamma$ mode.

In section 2 we point out some features of the process $\gamma \gamma
\longrightarrow \chi^+ \chi^- \gamma$ in the two-photon centre-of-mass
frame. The more realistic case of a laser back-scattering experiment,
and possible ways of eliminating backgrounds, are taken up in section
3. We summarise our numerical results and conclude in section 4.

\section{$\gamma \gamma \longrightarrow \chi^+ \chi^- \gamma$
        for monochromatic photon beams}
      
\begin{figure}[htb]
\vspace*{-3.5cm}
\centerline{\hspace*{-3em}
\epsfxsize=9.0cm\epsfysize=11.0cm
                     \epsfbox{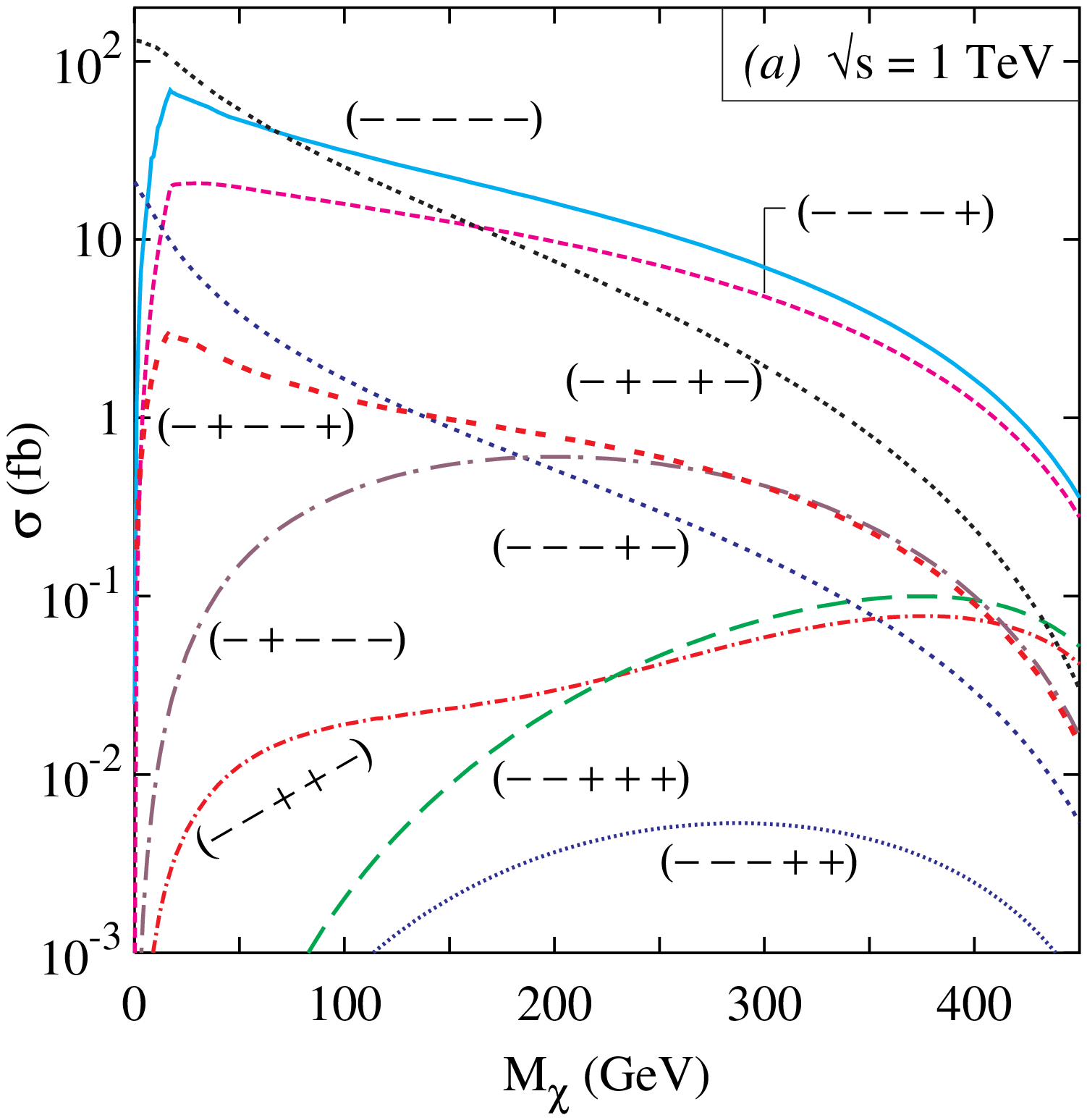}
\epsfxsize=8.0cm\epsfysize=10.0cm
                     \epsfbox{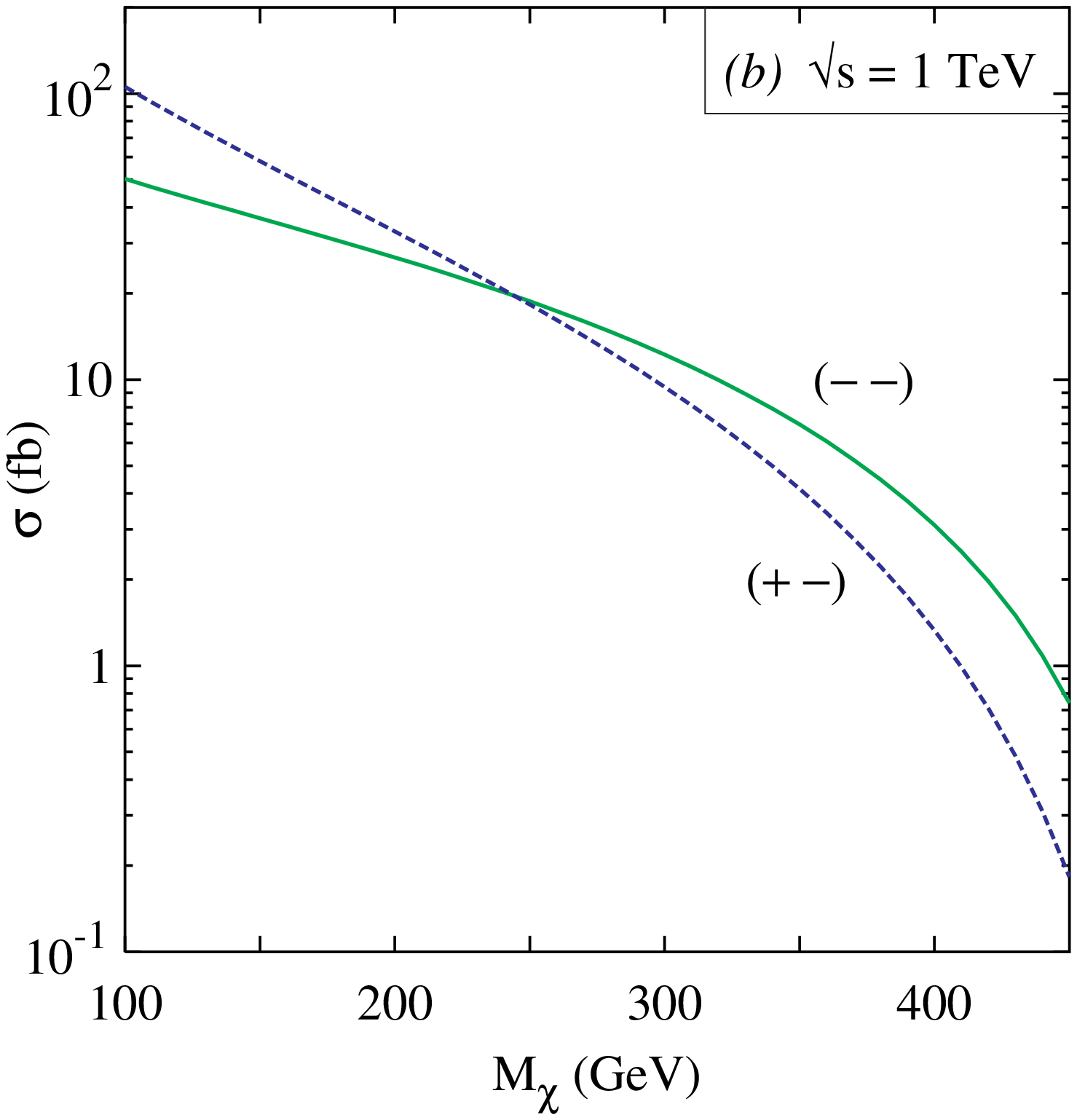}
}
\caption{\em Cross-section for 
  $\gamma \gamma \rightarrow \chi^+ \chi^- \gamma$ for monochromatic
  photon beams as a function of the chargino mass.  Panel (a) shows
  the cross-sections for each of the nine independent helicity
  channels, while panel (b) gives the two independent combinations
  when the final state polarisations have been summed over. The cuts
  applied are as in the text.  }
\label{fig:monochrome}
\end{figure}

      The production, governed solely by quantum electrodynamics,
      proceeds through six Feynman diagrams.  To understand the signal
      profile, it is useful to consider the individual contributions
      from each of the various helicity combinations. Clearly, not all
      of the 32 possible amplitudes are independent, related as they
      are by discrete symmetries (charge conjugation and parity). With
\be
    \sigma_{ijkln} \equiv \sigma(\gamma_i \gamma_j \rightarrow 
                                \chi^+_k \chi^-_l \gamma_n)
\ee
where the subscripts (taking values $\pm$) refer to the respective
particle helicities, we have
\be
\ba{rcl c l c l c l}
\sigma_{-----} & = & \sigma_{+++++} & & & & & & \\
\sigma_{----+} & = & \sigma_{++++-} & & & & & & \\
\sigma_{---+-} & = & \sigma_{+++-+} &= &\sigma_{--+--} &= & \sigma_{++-++} 
        & & \\
\sigma_{---++} & = & \sigma_{+++--} &= &\sigma_{--+-+} &= & \sigma_{++-+-} 
        & & \\
\sigma_{--++-} & = & \sigma_{++--+} & & & & & & \\
\sigma_{--+++} & = & \sigma_{++---} & & & & & & \\
\sigma_{-+---} & = & \sigma_{+-+++} &= &\sigma_{-++++} &= & \sigma_{+----} 
        & & \\
\sigma_{-+--+} & = & \sigma_{+-++-} &= &\sigma_{-+++-} &= & \sigma_{+---+} 
        & & \\
\sigma_{-+-+-} & = & \sigma_{+-+-+} &= &\sigma_{-++-+} &= & \sigma_{+--+-} 
        & & \\
               & = & \sigma_{-++--} & = & \sigma_{+--++} &= &
                        \sigma_{-+-++} &= & \sigma_{+-+--}. 
\ea
        \label{helicity_comb}
\ee
The above relations are true not only for the total cross-sections but
also for any partial sum, as long as the two charginos are subjected
to identical phase space constraints. Thus we have chosen to show, in
Fig.~\ref{fig:monochrome}($a$), the cross-sections corresponding to
the nine independent helicity combinations appearing in the first
column of each of the above equations.

The following observations are in order here. 
\begin{itemize} 
\item Each of the cross-sections is beset with kinematical
  singularities.  One of them corresponds to the final state photon
  being a soft one.  The other corresponds to mass and collinear
  singularities in the event of a vanishing chargino mass.
\item The electromagnetic vertex is helicity preserving. As all the
  final states with identical polarisation states for the chargino
  pair must have encountered a helicity flip, the corresponding
  amplitudes must be proportional to at least $M_\chi$. Thus, in the
  limit of vanishing chargino mass, these particular amplitudes should
  approach zero.
\end{itemize}
Clearly, wherever applicable, the two effects mentioned above pull the
cross-sections in different directions, one enhancing it, the other
suppressing.  However, if we impose a minimal energy requirement on
the final state particles as also demand that they be (at least
slightly) away from the beam pipe and also not collinear with each
other, then the kinematic singularities mentioned above are no longer
present in the partial cross-section.  In such an event, the helicity
reversal argument, wherever applicable, would prevail and pull the
cross-section down for sufficiently small chargino mass.
Figure~\ref{fig:monochrome}($a$) confirms this.  The cuts applied here
are $2^{\circ} \leq \theta_{\gamma}, \theta_{\chi^{\pm}} \leq
178^{\circ}$ and $\theta_{\chi^{\pm} \gamma} > 5^{\circ}$, where
$\theta_{\chi^{\pm} \gamma}$ is the angular separation of the photon
with the charginos.
%

It is instructive to examine Fig.~\ref{fig:monochrome}($a$) in some
detail.  The fall-off at large chargino masses ($M_\chi \rightarrow
\sqrt{s_{\gamma \gamma}} / 2$) is but a reflection of phase space
suppression.  In the small $M_\chi$ regime, the rapid fall in
$\sigma_{ij--n}$ is a consequence of the aforementioned helicity-flip
in the corresponding amplitudes.  Similarly the continued growth at
small masses for $\sigma_{-+-+-}$ and $\sigma_{---+-}$ is symptomatic
of the collinear singularity. The case of $\sigma_{---++}$ might seem
counterintuitive. However, this particular amplitude is essentially
governed by a double-helicity flip with the consequent $M_\chi^4$
behaviour of the cross-section for small masses.  Some additional
insight can be obtained if one considers the partial wave
decomposition of the amplitudes into different angular momentum
states. For example, the $J = 2$ initial state prefers to go to a
final state where the chargino pair can be thought of to be in $J = 1$
state and hence the final state charginos have opposite polarisations
(in other words, $\sigma_{- + - + - } = \sigma_{- + - + + }\gg
\sigma_{- + - - \pm}$).  Similar arguments can be constructed for the
other helicity amplitudes as well.

While the discussion above helps us in understanding the dynamics of
the process under consideration, the helicity of the photon in the
final state is virtually immeasurable. Furthermore, since we would be
primarily interested in the case of the chargino decaying into a
neutralino and a pion, the chargino polarisation information is also
lost.\footnote{In fact, even if we were to consider fermionic decay
  modes such as $\chi^+ \rightarrow \chi^0 \ell^+ \nu$, the effects of
  chargino polarisation are essentially averaged over as long as the
  decay products are invisible.}  Thus, we might as well sum over the
final state polarisations, thereby reducing the number of independent
cases to only two:
\be
\ba{rclcl}
        \sigma_{--} & \equiv & \dis 
                        \sum_{k,l,n} \sigma_{--kln} & = & \sigma_{++} \\
        \sigma_{-+} & \equiv & \dis 
                        \sum_{k,l,n} \sigma_{-+kln} & = & \sigma_{+-} 
\ea
\ee
In Fig.~\ref{fig:monochrome}($b$), we explicitly demonstrate the
$M_\chi$-dependence of these cross-sections, again for the ideal case
of monochromatic beams. The cuts applied in drawing the figures are
$E_{\gamma}>5$ GeV and $5^{\circ}\leq\theta_\gamma\leq 175^{\circ}$.
Note that $\sigma_{+-}$ dominates for small values of $M_\chi$ ceding
way to $\sigma_{--}$ at large chargino masses. The turnover point 
could also be inferred from a study of Fig.~\ref{fig:monochrome}($a$).
The exact location approaches $M_\chi \sim \sqrt{s_{\gamma \gamma}} /
4$ from below as the energy increases.

\section{Signals from backscattered photons and background elimination}

While an almost 100\% polarised photon beam is possible, obtaining an
intense high energy monochromatic beam is a near impossibility. In an
actual experiment, one proposes to scatter laser beams off the
$e^{+}e^{-}$ or $e^{-}e^{-}$ pair in the linear collider, and make the
scattered photons collide against each other. The cross-section for a
subprocess with a given centre-of-mass energy has to be folded with
the energy distributions of both the photons, which are essentially
Compton spectra. Not only has this spectrum a spread in the photon
energy, it is not even in a pure polarisation state. The exact shape
of the spectrum and the polarisation density matrix is determined by
the polarisation of the initial laser beam and the electron
(positron).

The basic parameters in the calculation are the energies of the
primary electron (positron) and the laser beam ($E_b$ and $E_l$
respectively), their polarisations, and the angle of
incidence ($\theta$) between the electron beam and the laser. The
quantity
\begin{equation}
z = {\frac{4E_b E_l}{m^2_e}}\cos^2 \frac{\theta}{ 2}
\end{equation}
determines the maximal fraction of the electron energy carried by the
scattered photon~\cite{telnov}.  An arbitrary increase in this
fraction results in pair production through the interaction of the
incident and scattered photons.  $z~=~2(1~+~\sqrt{2})$ is considered
to be an optimal choice~\cite{telnov} in this respect, and such a
value of $z$ has been adopted in this calculation. Expectedly, the
said cross-sections are beset with kinematical singularities and hence
are well-defined only when regulated (in other words, when phase space
constraints are imposed).

Let us start by focusing on the invisible decay
\be
        \chi^\pm \rightarrow \chi^0 + \pi^\pm
                \label{decay}
\ee
when pion is expected to be soft. 
Thus, the signal is 
\be
        \gamma \gamma  \rightarrow 
                \gamma + \mbox{\rm missing\ energy-momentum}.
        \label{signal}
\ee
To ensure that such a photon is visible, we demand that it is emitted
sufficiently away from the beam pipe and that it carries sufficiently
large transverse momentum. To be specific, the events are subjected to
the following cuts:
\be
\ba{rclcl}
  10^\circ &\le &\theta_{\gamma} & \le & 170^\circ \\ 
  \not{p_T} & = & p_T (\gamma) & \ge & 10 \ {\rm GeV}\\
  & & E(\gamma) & < & 100 \ {\rm GeV}  \\
  & & E(\pi^\pm) & < & 5 \ {\rm GeV}  
        \label{cuts}
\ea
\ee
where the last condition has been put to ensure invisibility of the
pions\footnote{Were we to consider the decay mode $\chi^+ \rightarrow
  \chi^0 \ell^+ \nu$, a similar restriction on the lepton energy would
  be imposed instead.}.  As will be seen below, these criteria not
only ensure visibility of the signals and and ward off singularities,
but also serve to eliminate practically all of the SM background. The
restriction on the maximal photon energy might seem puzzling at this
moment, but the need to impose such a requirement would become clearer
as we proceed.

Armed with the above requirements, we can convolute the cross-section
with the photon spectra. As has been mentioned, the latter are
determined by the polarisations of the incident laser and the
electron-positron pair. Clearly, 16 such distinct combinations are
possible. However, with the final state polarisations having been
summed over, it can easily be seen that only 6 of these combinations
are independent. Rather than consider all six, we shall focus our
attention on the two particular combinations (apart from the simplest
case, viz., the unpolarised one), that result in the largest
cross-sections for relatively heavy charginos.
\begin{figure}[htb]
\vspace*{-1in}
\centerline{
\epsfxsize=8.0cm\epsfysize=10.0cm
                     \epsfbox{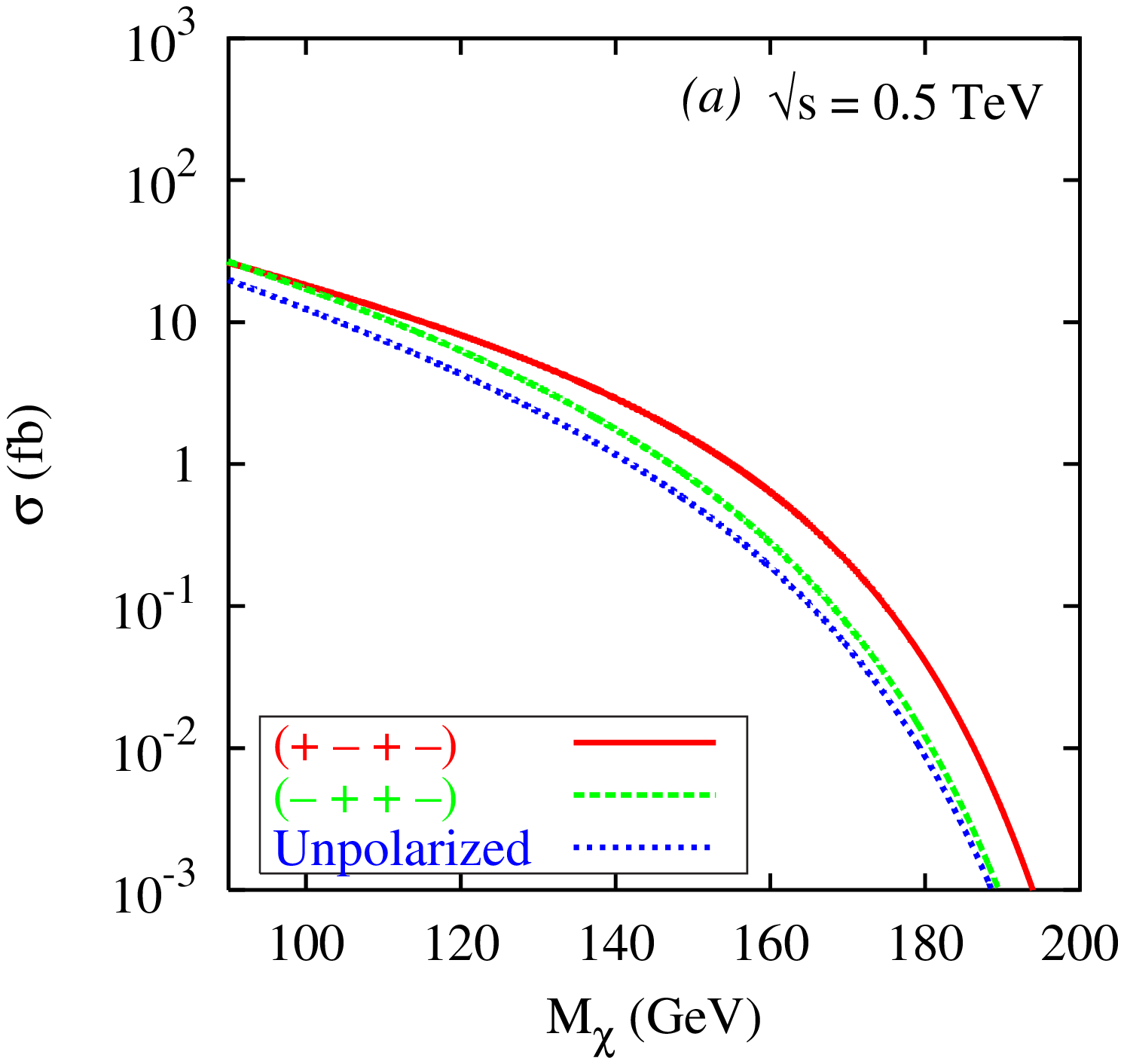}
\epsfxsize=8.0cm\epsfysize=10.0cm
                     \epsfbox{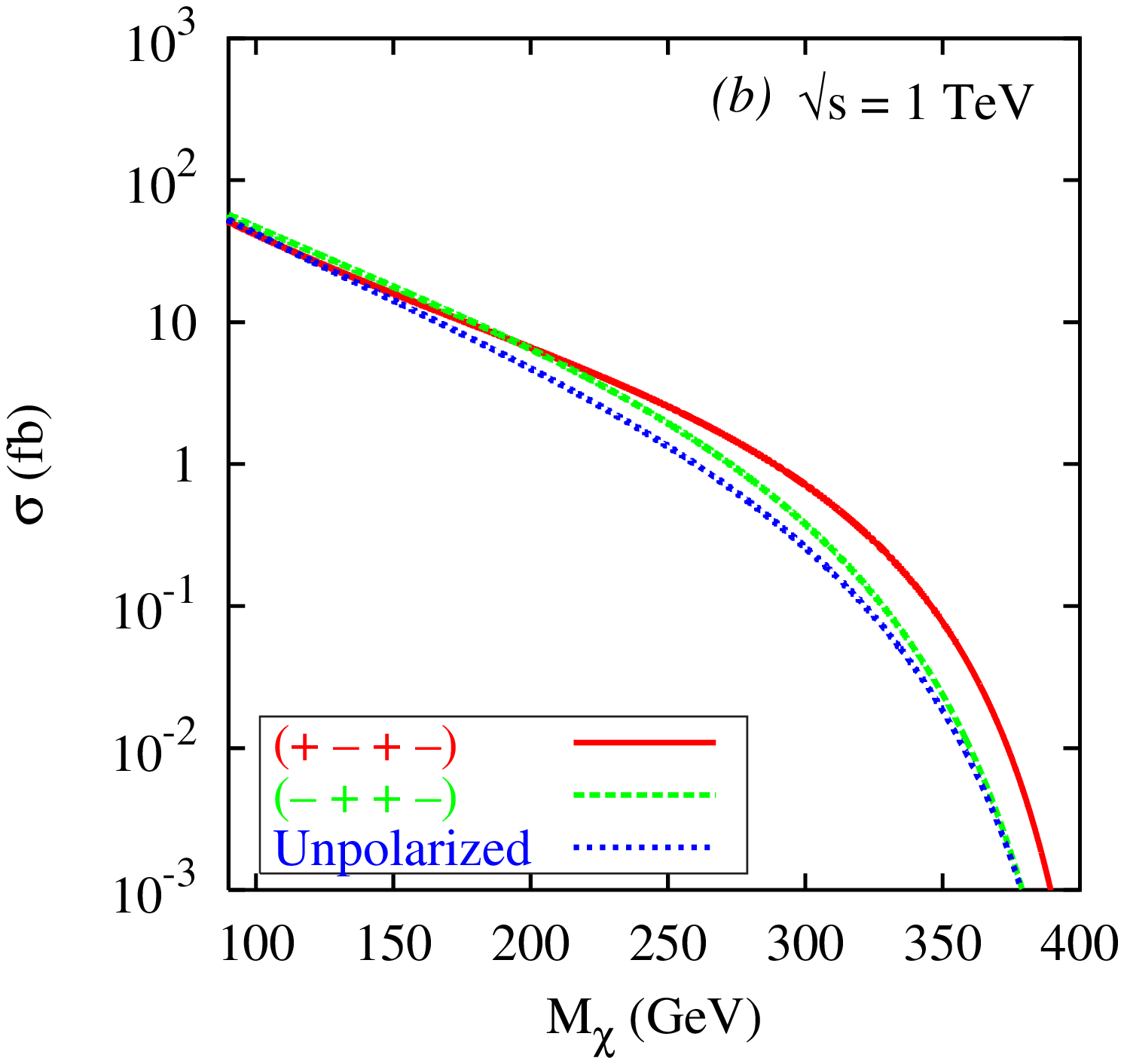}
}
\caption{\em Signal cross-sections as a function of chargino mass
  for two specific electron-electron centre-of-mass energies.  The
  cuts of Eqn.~(\protect\ref{cuts}) have been imposed.  The
  parenthetic combinations reflect the initial polarisations:
  (electron$_1$, laser$_1$, electron$_2$, laser$_2$) with 100\%
  polarisations for the lasers and 90\% for the electrons.  }
\label{fig:signal}
\end{figure}

Figure~\ref{fig:signal} shows the final cross-sections after applying
the above set of cuts, for two values of the electron-positron (or
electron-electron) centre-of-mass energy.  Polarisation efficiencies
of $90\%$ for electrons and $100\%$ for photons have been assumed.  As
the graphs show, the cross-sections have a significant dependence on
the beam polarisation choice for the entire range of chargino masses
considered. In particular, for large $M_\chi$, the dependence is quite
sizable with the different cross-sections varying by almost an order
of magnitude.

While we postpone comments on the numerical results till the next
section, it is essential at this stage to enter into a discussion on
the sources of backgrounds.  The SM backgrounds to the signal of
Eqn.~(\ref{signal}) arise from each of the following processes:
\begin{itemize}
\item $\gamma \gamma \rightarrow {\cal B}^+ {\cal B}^- \gamma$, 
        where ${\cal B}$ is a 
      light charged boson ($\pi$, $K$, $\rho$ {\em etc.});
\item $\gamma \gamma \rightarrow f \bar f \gamma$ (where $f$ is a lepton
      or a light quark) with the leptons escaping detection;
\item  $\gamma \gamma \rightarrow W^+ W^- \gamma$ with the $W$ decay 
        products escaping detection;
\item $\gamma \gamma \rightarrow \gamma Z$ with the $Z$ decaying 
        into neutrinos.
\end{itemize}
We  consider each in turn.

\underline{$\gamma \gamma \rightarrow {\cal B}^+ {\cal B}^- \gamma$}:
Events such $\pi^+ \pi^- \gamma$ production could be approximately
studied assuming naive scalar electrodynamics. While this is expected
to work fairly well at low momentum transfers (where the pion
structure is not resolved), one would be justified in questioning the
applicability for high energy processes such as the one we are
concerned with. Notwithstanding this criticism, this method is
expected to yield at least an order of magnitude estimate of the
rates.  Using this approach, we find that the admittedly large
differential cross-section is heavily biased towards very energetic
pions.  Once the criteria of Eqn.~(\ref{cuts}) are imposed, this
source of backgrounds is essentially eliminated. Analogous statements
hold for other possible mesons and hadrons with the comprehensive
result that this background is negligible.

\underline{$\gamma \gamma \rightarrow f \bar f \gamma$}: In order to
fake our signal, the pair of charged particles have to travel close
enough to the beam axis so as to escape detection. The exact criteria
for this to happen are, of course, dependent on the (as yet unknown)
detector design.  However, it is reasonable to impose the following
event detection criteria~\cite{chen_drees}.
\begin{itemize}
\item Any charged fermion emitted at an angle greater than $10^\circ$
  with the beam axis is detected as long as its energy is at least 5
  GeV.
\item Instrumentation in the region between $10^\circ$ and $1.2^\circ$
  with the beam axis would make detection possible there, provided the
  particle has a minimum energy of 50 GeV.
\item At angles less than $1.2^\circ$, no detection would be possible.
\end{itemize}
Any particle that does not satisfy at least one of the above criteria
would then escape undetected and hence contribute to the missing
energy-momentum. Explicit calculation shows that no such event
satisfies the event selection criteria of Eqn.~(\ref{cuts}) and hence
the corresponding background is eliminated completely.

\underline{$\gamma \gamma \rightarrow W^+ W^- \gamma$}: Quite
analogous to the previous case, this particular process would
contribute only if the decay products of the $W$'s escape detection.
It is quite obvious that this particular background would be even
smaller than the $f \bar f \gamma$ one, and again of no concern.

\begin{figure}[htb]
\vspace*{-1.8in}
\centerline{
\epsfxsize=10.0cm\epsfysize=14.0cm
                     \epsfbox{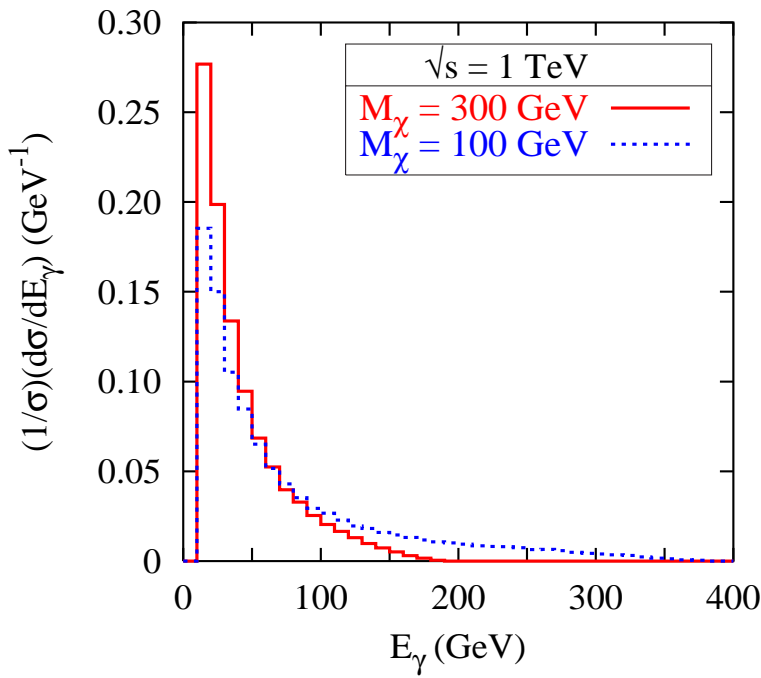}
}
\caption{\em The energy distribution for the outgoing photon 
        for the signal process. Two different chargino masses 
        are considered.
        The cuts of Eqn.~(\protect\ref{cuts}) have been imposed.
 }
\label{fig:energy_distrib}
\end{figure}

\underline{$\gamma \gamma \rightarrow \gamma Z$}: Interestingly, this
one-loop process proves to be the largest background. The
corresponding cross-section, calculated in~\cite{jikia}, is of the
order of 30~(50) $fb$ for $\sqrt{s_{e^+ e^-}} = 500~(1000)\ {\rm
  GeV}$.  Folding in the invisible decay width of the $Z$, this would
imply an ``irreducible'' background of 6~(10) $fb$. A naive comparison
with Figs.~\ref{fig:signal} thus suggests a severely limited mass
reach of our signal.  Fortunately, event kinematics comes to our
rescue. Thinking, for the moment, in terms of monochromatic photon
beams, the outgoing photon, too, would be monochromatic\footnote{The
  energy spread due to a possibly off-shell $Z$ is negligible.}  with
an energy of $E_\gamma = (s_{\gamma \gamma} - m_Z^2) / 2
\sqrt{s_{\gamma \gamma}}$.  On the other hand, the photon in the
signal process has a wide energy distribution peaking, typically, at
smaller $E_\gamma$. In the ideal case, then, eliminating photons lying
within a relatively narrow energy band would have solved our problem.
In reality though, one has to convolute this result with the photon
spectrum. Two factors come into play here: ($i$) the cross-section
$\sigma(\gamma \gamma \rightarrow \gamma Z)$ falls off very fast below
$\sqrt{s_{\gamma \gamma}} \lsim 170 \ {\rm GeV}$; and ($ii$) with our
favoured choice of beam polarisations (namely, opposite polarisations
for incoming electron and laser), the photon beam is strongly peaked
at large momentum fractions. Together, these imply that, even on
folding with the spectrum, the bulk of the contribution arises from
very energetic photons leading to typically large values of the
outgoing photon energy. The signal profile, on the other hand, is
strongly peaked at small outgoing photon energies (see
Fig.~\ref{fig:energy_distrib}).  That the spread is wider for smaller
chargino masses is, of course, easily understood. However, since the
cross-section for a smaller $M_\chi$ is larger, a larger fractional
loss in the signal is still not too costly.  More importantly, as
Fig.~\ref{fig:energy_distrib} shows, a restriction of $E_\gamma < 100
\ {\rm GeV}$ (see Eqn.~\ref{cuts}) leads to a very small loss of
signal, while suppressing the $\gamma Z$ background to a small
fraction of a femtobarn\footnote{Note, though, that our argument does
  not hold for $\gamma Z$ production starting from resolved photons.
  However, as Ref.\protect\cite{jikia} points out, the corresponding
  cross-sections by themselves are more than a magnitude
  smaller.}.  To summarise, we have established that the selection
criteria of Eqn.~(\ref{cuts}) serve to make our signal almost
background-free.

\section{Numerical results and conclusions}

In the absence of any background, 5 events would constitute a
discovery. A perusal of Figures~\ref{fig:signal} (along with
reasonable estimates for detector efficiencies) thus gives us the
discovery potential for a particular choice of beam polarisation.
Assuming an integrated luminosity of 100 $fb^{-1}$ , an invisible
chargino of mass upto about 165--175 GeV can thus be detected at a
linear collider operating at $\sqrt{s}~=~500$~GeV.  For a 1 TeV
collider, the limit can easily go up to 370 GeV or so\footnote{The
  kinematic limit for the two modes are 207 GeV and 414 GeV
  respectively.}.  This should be contrasted with the search limits
that can be reached through radiative chargino production in
electron-positron collision. Although it has been claimed that
charginos can be probed upto the kinematic limit in the latter case,
such a claim is tenable only when the sneutrino-induced diagrams do
not contribute appreciably to the production rate. As has been shown
in Ref.~\cite{single_photon}, the signal rate suffers a reduction by
nearly an order of magnitude due to destructive interference when
sneutrino masses are close to the lower limit obtained from LEP. Thus,
an unambiguous exploration of the chargino-neutralino sector is not
possible from $e^{+}e^{-}$ collision data.  A further consideration is
that of the $\nu\bar{\nu}\gamma$ background. While the part that
emanates from $Z\gamma$ background can be largely eliminated by
registering events within a certain window in the photon energy, the
non-resonant part needs a much more careful reconsideration in terms of
photon energy and momentum~\footnote{Note, however that right
  polarising the electron beam could suppress background while
  eliminating the aforementioned model-dependence of the signal.}.
The errors involved in such reconstruction, and the efficiency factor
coming therein, may considerably reduce the efficacy of invisible
chargino searches in this channel. Signals suggested in $\gamma\gamma$
collision can be advantageous in this respect, thanks to both the
total absence of the $\nu \bar{\nu} \gamma$ backgrounds as well as the
lack of interference, in the signal cross-section, from the
sneutrino-induced channels.

Another matter of particular interest is the determination of the
chargino mass. Unlike in the case of an $e^{+}e^{-}$ collider, the
cross-section here is a function of only $M_\chi$ and the
centre-of-mass energy. As Figs.~\ref{fig:signal} show, the dependence
is quite a marked one, especially for the ($-++-$) polarisation
combination. Thus, event counting itself would allow a fairly good
determination of $M_\chi$, provided sufficient luminosity is
available. Furthermore, the difference in the functional dependence of
$\sigma(M_\chi)$ for different polarisation choices could be exploited
for consistency checks. In addition to this, one would expect the edge
of the photon energy spectrum to give an independent measurement.
However, as this is true in principle, in practice this turns out to
be of little use.  Even for an integrated luminosity of $100~{\rm
  fb}^{-1}$, the event rates fall to such low values near the edge
that the corresponding errors would be too large for this method to be
useful.

So far, we have assumed that $\chi_1^{\pm}$ decay to $\chi_1^{0}
\pi^\pm$ has $100 \%$ branching ratio. This, however, is a model
dependent feature. The width in this channel primarily depends on the
mass difference ($\Delta m$) between $\chi_1^{0}$ and $\chi_1^{\pm}$.
In anomaly mediated models, $\Delta m$ cannot exceed a few hundreds of
MeV. In that case, $\chi_1^{0}\pi^\pm$ is the overwhelmingly dominant
channel in which the chargino can decay. Apart from AMSB, almost
invisible charginos can occur for $\mu \gg M_{1,2}$ (with $M_2$
substantially smaller than $M_1$) or $\mu \ll M_{1,2}$. The first
choice is very similar to the AMSB scenario, where the LSP (as well as
the lighter chargino) is basically a wino. In the second case
(Higgsino LSP), $\Delta m$ can be relatively larger so that the
three-body decay of a $\chi_1^{\pm}$ leading to
$l^\pm\,\nu\,\chi_1^{0}$ can take place. The opening up of such a mode
reduces the branching ratio of the two-body channel down to $5 - 15
\%$. Invisibility of the lighter chargino in such cases is dictated by
the softness of the lepton coming from the three-body channel as well
as the pion from the two-body channel.  The invisible branching
fraction of $\chi_1^{\pm}$ for such invisible decays can then vary
from $33 - 40 \%$ depending on the parameters.  With such branching
ratios, the signals discussed here enable us to probe a lighter
chargino mass upto about $175~(340)$~GeV with $e^{+}e^{-}$
centre-of-mass energy of $0.5~(1)$~TeV and integrated luminosity of
$100~fb^{-1}$. If $\Delta m < m_\pi$, then $l^\pm\,\nu\,\chi_1^{0}$ is
the only possible decay mode of the $\chi_1^{\pm}$. Here the leptons
will be so soft that we can expect a signal of the type (single photon
+ missing energy) for this case as well.

Before we conclude, it may be relevant to note that ways of probing an
invisibly decaying chargino have also been suggested in the context of
the LHC. The proposed methods consist in either the detailed study of
superparticle cascades and the various
$jets~+~leptons~+~missing~energy$ final states~\cite{tata}, or the
analysis of $forward~jets~+~missing~energy$ signals~\cite{vbf} induced
by gauge boson fusion into chargino pairs. The search limit for
invisible charginos in the first case, however, is again crucially
dependent on the slepton (sneutrino) mass parameters. In the second
method, the search limits are independent of this parameter, and can
extend upto chargino masses of 300--500 GeV at various confidence
levels. However, backgrounds cannot be completely removed there, and
the success of the analysis depends on the precision of the background
estimates. The $\gamma\gamma$ channel of invisible chargino
production, being largely background-free with the suggested event
selection criteria, can thus be complementary to the searches carried
out at a hadron collider.

{\bf Acknowledgement:} D.C. thanks the Department of Science and
Technology, India for financial assistance under the Swarnajayanti
Fellowship grant.  B.M. acknowledges partial support from the Board of
Research in Nuclear Sciences, Government of India. A.D. thanks Academy
of Finland (project number 48787) for financial support.

\newcommand{\plb}[3]{{Phys. Lett.} {\bf B#1} (#3) #2}                  %
\newcommand{\prl}[3]{Phys. Rev. Lett. {\bf #1} (#3) #2}        %
\newcommand{\rmp}[3]{Rev. Mod.  Phys. {\bf #1} (#3) #2}             %
\newcommand{\prep}[3]{Phys. Rep. {\bf #1} (#3) #2}                     %
\newcommand{\rpp}[3]{Rep. Prog. Phys. {\bf #1} (#3) #2}             %
\newcommand{\prd}[3]{Phys. Rev. {\bf D#1} (#3) #2}                    %
\newcommand{\prc}[3]{{Phys. Rev.}{\bf C#1} (#3) #2}  
\newcommand{\jhep}[3]{{Jour. High Energy Phys.\/} {\bf #1} (#3) #2}%
\newcommand{\np}[3]{Nucl. Phys. {\bf B#1} (#3) #2}                     %
\newcommand{\npbps}[3]{Nucl. Phys. B (Proc. Suppl.) 
           {\bf #1} (#3) #2}                                           %
\newcommand{\sci}[3]{Science {\bf #1} (#3) #2}                 %
\newcommand{\zp}[3]{Z.~Phys. C{\bf#1} (#3) #2}                 %
\newcommand{\mpla}[3]{Mod. Phys. Lett. {\bf A#1} (#3) #2}             %
 \newcommand{\apj}[3]{ Astrophys. J.\/ {\bf #1} (#3) #2}       %
\newcommand{\astropp}[3]{Astropart. Phys. {\bf #1} (#3) #2}            %
\newcommand{\ib}[3]{{ ibid.\/} {\bf #1} (#3) #2}                    %
\newcommand{\nat}[3]{Nature (London) {\bf #1} (#3) #2}         %
 \newcommand{\app}[3]{{ Acta Phys. Polon.   B\/}{\bf #1} (#3) #2}%
\newcommand{\nuovocim}[3]{Nuovo Cim. {\bf C#1} (#3) #2}         %
\newcommand{\yadfiz}[4]{Yad. Fiz. {\bf #1} (#3) #2;             %
Sov. J. Nucl.  Phys. {\bf #1} #3 (#4)]}               %
\newcommand{\jetp}[6]{{Zh. Eksp. Teor. Fiz.\/} {\bf #1} (#3) #2;
           {JETP } {\bf #4} (#6) #5}%
\newcommand{\philt}[3]{Phil. Trans. Roy. Soc. London A {\bf #1} #2  
        (#3)}                                                          %
\newcommand{\hepph}[1]{(electronic archive:     hep--ph/#1)}           %
\newcommand{\hepex}[1]{(electronic archive:     hep--ex/#1)}           %
\newcommand{\astro}[1]{(electronic archive:     astro--ph/#1)}         %

\vspace*{2ex}


\end{document}